\documentclass{emulateapj}
\usepackage{apjfonts}

\newcommand{\ddt}[1]{\frac{d #1}{d t}}
\newcommand{\zh}{Z_{\mathrm{H}}}
\newcommand{\sh}{\Sigma_{\mathrm{H}}}
\newcommand{\she}{\Sigma_{\mathrm{He}}}
\newcommand{\estarh}{E^{*}_{\mathrm{H}}}
\newcommand{\estarhe}{E^{*}_{\mathrm{He}}}
\newcommand{\gh}{\gamma_{\mathrm{H}}}
\newcommand{\ghe}{\gamma_{\mathrm{He}}}
\newcommand{\epsh}{\epsilon_{\mathrm{H}}}
\newcommand{\epshe}{\epsilon_{\mathrm{He}}}
\newcommand{\epscool}{\epsilon_{\mathrm{cool}}}
\newcommand{\xout}{X_{0}}
\newcommand{\yout}{Y_{0}}
\newcommand{\zout}{Z_{0}}
\newcommand{\dshdt}{\frac{\partial \sh}{\partial t}}
\newcommand{\dshedt}{\frac{\partial \she}{\partial t}}
\newcommand{\dzhdt}{\frac{\partial Z_{\mathrm{H}}}{\partial t}}

\newcommand{\sdot}{\dot{\Sigma}}
\newcommand{\dsig}{d \Sigma}
\newcommand{\thy}{T_{\mathrm{H}}}
\newcommand{\thel}{T_{\mathrm{He}}}
\newcommand{\fhy}{F_{\mathrm{H}}}
\newcommand{\fhe}{F_{\mathrm{He}}}

\newcommand{\dthdt}{\frac{\partial T_{\mathrm{H}}}{\partial t}}

\newcommand{\cp}{C_{\mathrm{p}}}

\newcommand{\themthf}{\thel^{4}-\thy^{4}}
\newcommand{\pd}[2]{\frac{\partial #1}{\partial #2}}
\newcommand{\eq}[1]{#1^{\mathrm{eq}}}
\newcommand{\delsig}{\Delta \Sigma}
\newcommand{\lacc}{l_{\mathrm{acc}}}
\newcommand{\llacc}{\log(\lacc)}
\newcommand{\tacc}{t_{\mathrm{acc}}}
\newcommand{\reltacc}{\Re ( \lambda ) \tacc}
\newcommand{\oH}{\omega_{\mathrm{H}}}
\newcommand{\tH}{t_{\mathrm{H}}}
\newcommand{\slayer}{\Sigma_{\mathrm{layer}}}

\begin{document}

\title{A Two-Zone Model for Type I X-ray Bursts on Accreting Neutron Stars}
\shorttitle{Two-Zone Burst Model}

\author{Randall L.\ Cooper and Ramesh Narayan}
\affil{Harvard-Smithsonian Center for Astrophysics, 60 Garden 
Street, Cambridge, MA 02138}

\email{rcooper@cfa.harvard.edu, rnarayan@cfa.harvard.edu}

\begin{abstract}

We construct a two-zone model to describe hydrogen and helium burning
on the surface of an accreting neutron star and use it to study the
triggering of type I X-ray bursts.  Although highly simplified, the
model reproduces all of the bursting regimes seen in the more complete
global linear stability analysis of \citet{NH03}, including the regime
of delayed mixed bursts.  The results are also consistent with
observations of type I X-ray bursts.  At low accretion rates $\dot{M}
/ \dot{M}_{\mathrm{Edd}}\lesssim 0.1 $, thermonuclear helium burning
via the well-known thin-shell thermal instability triggers bursts.  As
$\dot{M}$ increases, however, the trigger mechanism evolves from the
fast thermal instability to a slowly growing overstability involving
both hydrogen and helium burning.  The competition between nuclear
heating via the $\beta$-limited CNO cycle and the triple-$\alpha$
process on the one hand, and radiative cooling via photon diffusion
and emission on the other hand, drives oscillations with a period
approximately equal to the hydrogen-burning timescale.  If these
oscillations grow, the gradually rising temperature at the base of the
helium layer eventually provokes a thin-shell thermal instability and
hence a delayed mixed burst.  This overstability closely resembles the
delayed mixed bursts of Narayan \& Heyl.  For
$\dot{M}/\dot{M}_{\mathrm{Edd}} \gtrsim 0.25$ there is no instability
or overstability, and there are no bursts.  Nearly all other
theoretical models, including detailed time-dependent multi-zone
calculations, predict that bursts should occur for all
$\dot{M}/\dot{M}_{\mathrm{Edd}} \lesssim 1$, in conflict with both our
results and observations.  We suggest that this discrepancy arises
from the assumed strength of the hot CNO cycle breakout reaction
$^{15}$O($\alpha$,$\gamma$)$^{19}$Ne in these other models.  That
observations agree much better with the results of Narayan \& Heyl and
our two-zone model, both of which neglect breakout reactions, may
imply that the true $^{15}$O($\alpha$,$\gamma$)$^{19}$Ne cross section
is much smaller than assumed in previous investigations.

\end{abstract} 


\section{Introduction}

Type I X-ray bursts are thermonuclear explosions that occur on the
surfaces of accreting neutron stars
\citep{Betal75,GH75,Getal76,BCE76,WT76,J77,MC77,LL77,LL78}.  They are
triggered by unstable helium or hydrogen burning \citep[for
reviews, see][]{LvPT93,LvPT95,C04,SB03}.  For a wide range of
accretion rates, the basic physics of the burst onset is that of the
well-known thin-shell thermal instability \citep{SH65,HvH75}, 
which operates as follows.  In
equilibrium, the energy that is generated by nuclear burning is
exactly balanced by the outward diffusion of photons coupled with
radiation from the stellar surface.  If the nuclear reaction
rates are sufficiently temperature-sensitive, a positive temperature
perturbation causes the nuclear energy generation rate to increase
faster than the rate at which the stellar surface can cool, causing a
thermonuclear runaway and hence a type I X-ray burst.  The physics of
this instability can be described quite well with a simple one-zone
model \citep{FHM81,P83,B98,CB00}.  If the amount of accreted matter
needed to ignite a burst were roughly independent of the accretion
rate $\dot{M}$, as suggested by simple models, one would expect the
recurrence time $t_{\mathrm{rec}}$ to vary as $\dot{M}^{-1}$.  This is
in reasonable agreement with observations for a range of $\dot{M}
\lesssim 0.1 \dot{M}_{\mathrm{Edd}}$, where $\dot{M}_{\mathrm{Edd}}$
denotes the mass accretion rate at which the accretion luminosity is
equal to the Eddington limit.  However, for $0.1 \lesssim \dot{M}/
\dot{M}_{\mathrm{Edd}} \lesssim 0.3$, observations show that 
$t_{\mathrm{rec}}$ increases with increasing $\dot{M}$, and bursts are 
found to cease
altogether for $\dot{M}/\dot{M}_{\mathrm{Edd}} \gtrsim 0.3$
\citep{vPCLJ79,vPPL88,Cetal03,RLCN06}.  Moreover, for $0.1 \lesssim
\dot{M}/ \dot{M}_{\mathrm{Edd}} \lesssim 0.3$, most of the accreted
matter appears to burn stably between consecutive bursts
\citep{vPPL88,intZetal03}.  Simple thin-shell thermal instability
models, on the other hand, predict that bursts should occur for all
$\dot{M}/\dot{M}_{\mathrm{Edd}} \lesssim 1$ and that 
$t_{\mathrm{rec}}$ should decrease monotonically with 
increasing $\dot{M}$.

\citet[][hereafter NH03]{NH03} developed a global linear stability
analysis of the accreted nuclear fuel on the surface of a neutron
star.  Unlike thin-shell thermal instability models, for which one
assumes that purely thermal perturbations trigger bursts, NH03's
analysis was quite general and made no assumptions as to the trigger
mechanism.  Apart from confirming the standard thermal instability for
$\dot{M} / \dot{M}_{\mathrm{Edd}} \lesssim 0.1$, NH03 discovered a new
regime of unstable nuclear burning for $0.1 \lesssim \dot{M}/
\dot{M}_{\mathrm{Edd}} \lesssim 0.3$ that they referred to as
``delayed mixed bursts.''  This regime reproduced the increase in
$t_{\mathrm{rec}}$ with increasing $\dot{M}$ seen in observations,
as well as the occurrence of considerable stable burning between
successive bursts.  It also predicted the absence of bursts for
$\dot{M}/ \dot{M}_{\mathrm{Edd}} \gtrsim 0.3$, again 
in agreement with observations.  Recently, \citet{CN05}
and \citet{CMSN06} have suggested that the stable burning that
precedes delayed mixed bursts is crucial for the production of the
carbon fuel needed to trigger superbursts \citep[for reviews of
superbursts, see][]{K04,C05}, since only stable nuclear burning can
produce carbon in sufficient quantities
\citep{SBCW99,Setal01,SBCO03,KHKF04,Wetal04,FBLTW05}.

It is perhaps not surprising that a global linear stability analysis
is able to reproduce type I X-ray burst observations better than
simple thin-shell thermal instability models.  However, the fact that
the results of all time-dependent multi-zone models with large
reaction networks conflict with observations as well
\citep{AJ82,TWL96,FHLT03,HCW05} is much more problematic.  Like the
global linear stability analysis, these latter models include no
preconceived instability criteria.  Therefore, they should in
principle be able to reproduce the regime of delayed mixed bursts and
the absence of bursts for $\dot{M}/ \dot{M}_{\mathrm{Edd}} \gtrsim
0.3$, as observed in nature.  The disagreement between the predictions
of time-dependent multi-zone burst models and the global linear
stability analysis of NH03 suggests that some crucial aspect of the
physics is different between the two.  Understanding this discrepancy
is of fundamental importance in the study of type I X-ray bursts.

While the global linear stability analysis of NH03, with its
identification of the delayed mixed burst regime, succeeded in
reproducing several puzzling observations of bursts, unfortunately,
the complexity of the model and the absence of a specific stability
criterion make it difficult to understand the basic physics of delayed
mixed bursts.  To rectify this situation, we construct here a simple
two-zone model that reproduces all of the basic features of the global
analysis.  We begin in \S \ref{themodel} with a derivation of our
model.  In \S \ref{eqandstab}, we discuss the equilibria of the model
and analyze their stability as a function of accretion rate.  We then
integrate the equations that govern the model and use the results in
\S \ref{tevolve} to explain the physics of the onset of delayed mixed
bursts.  We discuss our results in \S \ref{discussion}, and we
conclude in \S \ref{conclusions}.

\section{The Model}\label{themodel}

We assume that matter accretes spherically onto a neutron star of
gravitational mass $M = 1.4 M_{\odot}$ and areal radius $R = 10 \,
\mathrm{km}$ at an accretion rate per unit area $\sdot$ as 
measured in the local frame of the accreted plasma.  We consider
all physical quantities to be functions of the column depth $\Sigma$,
which we define as the rest mass of the accreted matter as measured
from the stellar surface divided by $4 \pi R^{2}$.  We denote the
Eulerian time and spatial derivatives as $\partial/\partial t$ and
$\partial/\partial \Sigma$, respectively, and we denote the Lagrangian
derivative following a parcel of matter as $D/Dt$, where $D/Dt =
\partial/\partial t + \sdot \partial/\partial \Sigma$.  We describe
the composition of the matter by the hydrogen mass fraction $X$,
helium mass fraction $Y$, and heavy element fraction $Z = 1-X-Y$.  The
mass fractions at $\Sigma=0$, $X_{0}$, $Y_{0}$, and $Z_{0}$, are those
of the accreted plasma.  In this work, we assume that all heavy
elements are CNO and that the composition of the accreted matter is
that of the Sun: $\xout = 0.7$, $Y_{0} = 0.28$, $\zout = 0.02$.

The following set of four partial differential equations governs the
thermal structure and time evolution of the material (e.g., NH03; 
see Table 1 of that paper for the definitions of the various
symbols):
\begin{equation}\label{transfereqn}
\frac{\partial T^{4}}{\partial \Sigma} = \frac{3\kappa F}{ac},
\end{equation}
\begin{equation}\label{entropyeqn}
\frac{\partial F}{\partial \Sigma} = T\frac{Ds}{Dt} - (\epsh+\epshe),
\end{equation}
\begin{equation}\label{dxdteqn}
\frac{DX}{Dt} = -\frac{\epsh}{\estarh},
\end{equation}
\begin{equation}\label{dzdteqn}
\frac{DZ}{Dt} = \frac{\epshe}{\estarhe}.
\end{equation}
Since we are interested in developing a simple model that avoids
unnecessary details, we follow \citet{B98} and set the opacity to a
constant value, $\kappa=0.4\kappa_{\mathrm{es}}=0.136$
$\mathrm{cm}^{2}\,\mathrm{g}^{-1}$, where $\kappa_{\mathrm{es}}$ is
the Thomson scattering opacity.  We assume that a nondegenerate, ideal
gas supplies the pressure $P$, which is a good approximation for the
shallow column depths we consider in this work.  The specific entropy
$s$ is in general a function of both $T$ and $P=g\Sigma$, and so
\begin{equation}\label{oldentropyeqn}
T\frac{Ds}{Dt} = \cp \frac{\partial T}{\partial t} + \cp \sdot \left (
\pd{T}{\Sigma} - \frac{T}{\Sigma} \nabla_{\mathrm{ad}} \right ),
\end{equation}
where we set the specific heat at constant pressure $\cp = 5
k_{\mathrm{B}}/2 m_{\mathrm{p}}$ for an assumed mean molecular weight
$\mu = 1$, and $\nabla_{\mathrm{ad}} \equiv (\partial \ln T/\partial
\ln \Sigma)_{s} = 2/5$ for an ideal gas.  We ignore the terms $\propto
\sdot$ for reasons we will give in \S \ref{eqandstab} \citep[this is a
standard approximation in most one-zone models of type I X-ray bursts,
e.g.,][]{B98,CB00} although we have done calculations with these terms
included, and we have verified that they are not important.  Thus,
equation (\ref{entropyeqn}) becomes
\begin{equation}\label{newentropyeqn}
\frac{\partial F}{\partial \Sigma} = \cp \pd{T}{t} - (\epsh+\epshe).
\end{equation}

The usual one-zone approach to type I X-ray bursts
\citep{FHM81,P83,B98,CB00} focuses primarily on helium burning, for
which a single zone is adequate.  Several authors have introduced more
elaborate two-zone models of type I X-ray bursts
\citep{BP79,BBL80,RL84,LR85,Y87}, but these models usually include
only helium burning, and all of the nuclear burning is confined to a
single zone.  \citet{JHI93} present a two-zone model of thermonuclear
burning on accreting white dwarfs that includes both hydrogen and
helium burning, but the two fuels are forced to burn in separate zones
by construction.  This approximation is too severe for realistic
modeling of type I X-ray bursts.  The global linear stability analysis
of NH03 clearly shows that the interaction between hydrogen and helium
burning plays a key role in delayed mixed bursts.  Motivated by this,
we describe in this paper a new model with the following two zones:
(i) a zone that begins at the surface of the star, where $\Sigma=0$,
and extends to the depth $\sh$ at which hydrogen is depleted via
nuclear burning, and (ii) a zone that begins at $\sh$ and extends to
the depth $\she$ at which helium is depleted via nuclear burning.
Hydrogen, helium, and CNO are all present in zone (i), while only
helium and CNO are present in zone (ii).  Furthermore, we include both
hydrogen and helium burning in zone (i), but we include only helium
burning in zone (ii) since hydrogen is not present.  Note that, in our
model, hydrogen fully depletes before all of the helium burns by
construction, so that $\she > \sh$.  This condition is satisfied over
the entire range of $\sdot$ of interest in this work.  We discuss the
case for which $\she < \sh$ in \S \ref{discussion}.  We make the
simplification that all thermonuclear processes within a zone take
place at the bottom of that zone.  Equation (\ref{entropyeqn}) thus
implies that the flux $F$ in a particular zone is a constant with
respect to $\Sigma$, and so $T^{4}$ varies linearly with $\Sigma$
within each zone (see equation \ref{transfereqn}).

\begin{figure}
\plotone{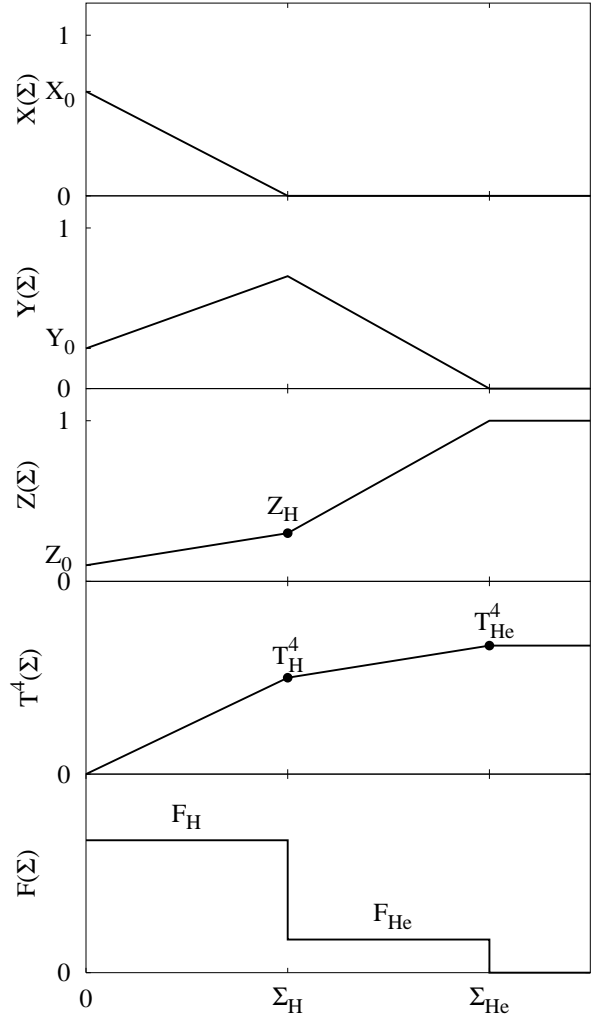}
\caption{A schematic representation of how the composition,
temperature, and flux vary as a function of column depth in the
proposed two-zone model.  $\xout$, $\yout$, and $\zout$ denote the
hydrogen, helium, and CNO mass fractions of the accreted plasma,
respectively.  Zone (i) extends from $\Sigma = 0$ to $\Sigma = \sh$,
and zone (ii) extends from $\Sigma = \sh$ to $\Sigma = \she$.}
\label{modelfig}
\end{figure}

In the simplest one-zone models, all parameters of the matter are
taken to be constant within the zone.  Clearly this prescription is a
gross simplification, for in reality the physical parameters $X$, $Y$,
$Z$, $T$, and $F$ vary smoothly as functions of $\Sigma$.  In order to
improve the accuracy of our model, we assume linear profiles with
respect to depth of the various quantities of interest (except $F$).
Thus we take
\begin{equation}\label{Xeqn}
X(\Sigma,t) = \left \{ 
\begin{array}{ll}
\xout (1-\Sigma/\sh), & \Sigma < \sh\\
0, & \Sigma > \sh,\\
\end{array} \right.
\end{equation}
\begin{equation}\label{Zeqn}
Z(\Sigma,t) =  \left \{ 
\begin{array}{lll}
\zout + (\zh-\zout)(\Sigma/\sh), & \Sigma < \sh\\
\zh + (1-\zh)(\Sigma-\sh)/\delsig, & \sh < \Sigma < \she\\
1, & \Sigma > \she,\\
\end{array} \right.
\end{equation}
\begin{equation}\label{Teqn}
T^{4}(\Sigma,t) =  \left \{ 
\begin{array}{lll}
\thy^{4} (\Sigma/\sh), & \Sigma < \sh\\
\thy^{4} + (\thel^{4}-\thy^{4}) (\Sigma-\sh)/\delsig, & \sh < \Sigma < \she\\
\thel^{4}, & \Sigma > \she,\\
\end{array} \right.
\end{equation}
\begin{equation}\label{Feqn}
F(\Sigma,t) =  \left \{ 
\begin{array}{lll}
\fhy, & \Sigma < \sh\\
\fhe, & \sh < \Sigma < \she\\
0, & \Sigma > \she,\\
\end{array} \right.
\end{equation}
where
\begin{equation}
\zh \equiv Z(\sh), \,\,\,\,\,\thy \equiv T(\sh), \,\,\,\,\,\thel
\equiv T(\she),
\end{equation}
\begin{equation}
\delsig \equiv \she - \sh,
\end{equation}
and the five quantities $\sh$, $\zh$, $\thy$, $\she$, and $\thel$ are
functions of time.  Note that equation (\ref{Teqn}) is exact for 
the aforementioned simplifications of constant opacity and constant 
flux within a zone.  Figure \ref{modelfig} shows a pictorial
representation of the model.  For simplicity, we assume that the flux
$F$ entering the bottom of the helium-burning zone from the stellar
interior is zero.  Integrating equation (\ref{transfereqn}) gives
\begin{equation}
\fhy = \frac{ac \thy^{4}}{3 \kappa \sh},
\end{equation}
\begin{equation}\label{fheeqn}
\fhe = \frac{ac (\thel^{4}-\thy^{4})}{3 \kappa \delsig}.
\end{equation}

For the relatively high accretion rates we consider in this work,
hydrogen burns to helium via the hot CNO cycle, so the corresponding
nuclear energy generation rate $\epsh = \estarh \gh Z$ is a function
only of $Z$, where $\gh = 9.1 \times 10^{-4}\,\mathrm{s}^{-1}$
\citep{HF65}.  Helium burns to carbon via the triple-$\alpha$
reaction.  The corresponding energy generation rate $\epshe$ is a
strong function of temperature, but it depends also on $Y$ and the
density $\rho$.  In this work, we simplify matters by approximating
$\epshe$ to be a function only of temperature \citep{HKT04}, so we set
\begin{equation}\label{epsheeqn}
\epshe(T) = \estarhe \ghe (T_{0}/T)^{3}\exp(-T_{0}/T),
\end{equation}
where $T_{0} = 4.4027\times10^{9}\,\mathrm{K}$ and we choose $\ghe =
0.1\,\mathrm{s}^{-1}$ such that $\epshe$ in our model is a reasonable
representation of the true $\epshe(T,Y,\rho)$ at densities typical of
those at which helium burns on accreting neutron stars.

We wish to derive a set of five differential equations that determine
the behavior of the five physical quantities $\sh$, $\zh$, $\thy$,
$\she$, and $\thel$.  To do this, we integrate equations
(\ref{dxdteqn}), (\ref{dzdteqn}), and (\ref{newentropyeqn}) from $0$
to $\sh$ and equations (\ref{dzdteqn}) and (\ref{newentropyeqn}) from
$\sh$ to $\she$.  This gives
\begin{equation}\label{xintegral}
\int_{0}^{\sh} \left ( \pd{X}{t} + \sdot \pd{X}{\Sigma} \right )\dsig
= -\int_{0}^{\sh} \frac{\epsh}{\estarh} \dsig,
\end{equation}
\begin{equation}\label{zintegral1}
\int_{0}^{\sh} \left ( \pd{Z}{t} + \sdot \pd{Z}{\Sigma} \right )\dsig
= \int_{0}^{\sh} \frac{\epshe}{\estarhe} \dsig,
\end{equation}
\begin{equation}\label{tintegral1}
\int^{\sh}_{0} \pd{F}{\Sigma} \dsig = \int^{\sh}_{0} \left ( \cp
\pd{T}{t} - \epsh - \epshe \right ) \dsig,
\end{equation}
\begin{equation}\label{zintegral2}
\int_{\sh}^{\she} \left ( \pd{Z}{t} + \sdot \pd{Z}{\Sigma} \right
)\dsig = \int_{\sh}^{\she} \frac{\epshe}{\estarhe} \dsig,
\end{equation}
\begin{equation}\label{tintegral2}
\int^{\she}_{\sh} \pd{F}{\Sigma} \dsig = \int^{\she}_{\sh} \left ( \cp
\pd{T}{t} - \epshe \right ) \dsig.
\end{equation}
Since $\epshe$ is a very strong function of temperature, most of the
energy generated via helium burning within a zone will be released
near the bottom of that zone, where the temperature is greatest.
Therefore, when we integrate over $\epshe$, we make the approximation
that the temperature throughout the zone is the temperature at the
base, so we set 
\begin{equation} \int^{\sh}_{0} \epshe \dsig \equiv
\epshe(\thy)\sh,
\end{equation}
\begin{equation}
\int^{\she}_{\sh} \epshe \dsig \equiv \epshe(\thel)\delsig.
\end{equation}
In contrast, $\epsh$ is a weak function of depth, and so the energy
generated via hydrogen burning within zone (i) will be released more
evenly throughout the zone.  Therefore, when we integrate over
$\epsh$, we perform the integral exactly according to equation
(\ref{Zeqn}):
\begin{equation}\label{epsheqn}
\int^{\sh}_{0} \epsh \dsig = \estarh \gh \int^{\sh}_{0} Z(\Sigma,t)
\dsig = \estarh \gh \left (\frac{\zh+\zout}{2} \right )\sh.
\end{equation}
To derive the system of differential equations that govern our model,
we evaluate the integrals in equations
(\ref{xintegral}-\ref{tintegral2}) according to the ans\"{a}tze of
equations (\ref{Xeqn}-\ref{Feqn}) and the approximations described
above.  Using the ans\"{a}tz for $X(\Sigma,t)$ in equation
(\ref{Xeqn}) and the result of equation (\ref{epsheqn}), equation
(\ref{xintegral}) becomes
\begin{eqnarray}
\int_{0}^{\sh} \pd{}{t}\left [\xout \left (1-\frac{\Sigma}{\sh} \right
) \right] \dsig + 
\sdot \int_{0}^{\sh}\pd{X(\Sigma,t)}{\Sigma} \dsig = \nonumber \\
-\gh \left (\frac{\zh+\zout}{2} \right )\sh.  \nonumber \\
\end{eqnarray}
From the fundamental theorem of calculus, and recalling that
$\sh=\sh(t)$, this gives
\begin{equation}
\xout \int_{0}^{\sh} \left ( \frac{\Sigma}{\sh^{2}} \pd{\sh}{t} \right
) \dsig + \sdot(0-\xout) = -\gh \left (\frac{\zh+\zout}{2} \right
)\sh.
\end{equation}

By performing analogous steps on equations
(\ref{zintegral1}-\ref{tintegral2}), equations
(\ref{xintegral}-\ref{tintegral2}) may now be written as
\begin{equation}\label{finalstep1}
\frac{1}{2} \xout \pd{\sh}{t} = \xout \sdot - \gh \left
(\frac{\zh+\zout}{2} \right ) \sh,
\end{equation}
\begin{equation}
\frac{1}{2} \left [\dzhdt \sh - (\zh-\zout)\dshdt \right ] =
\frac{\epshe(\thy)}{\estarhe} \sh - (\zh-\zout)\sdot,
\end{equation}
\begin{eqnarray}
\cp \left (\dthdt - \frac{\thy}{4 \sh}\dshdt \right ) \left
(\frac{4}{5} \sh \right ) = \estarh \gh \left (\frac{\zh+\zout}{2}
\right )\sh +  \nonumber \\  \nonumber \\ \epshe(\thy)\sh - (\fhy - \fhe), 
\nonumber \\
\end{eqnarray}
\begin{eqnarray}
\frac{1}{2} \left [\dzhdt \delsig - (1-\zh) \left (\dshedt+\dshdt
\right )\right ] =  \nonumber \\  \nonumber \\ \frac{\epshe(\thel)}{\estarhe} \delsig -
(1-\zh)\sdot,
\end{eqnarray}
\begin{equation}\label{finalstep5}
\cp \int^{\she}_{\sh} \pd{T(\Sigma,t)}{t}\dsig = \epshe(\thel)\delsig
- \fhe.
\end{equation}
As the quantities $\sh$, $\zh$, $\thy$, $\she$, and $\thel$ are
functions of only one independent variable $t$, equations
(\ref{finalstep1}-\ref{finalstep5}) can be rewritten as the following
set of coupled ordinary differential equations:
\begin{equation}\label{dshdteqn}
\ddt{\sh} = 2 \sdot - \frac{\gh \sh}{\xout}(\zh+\zout),
\end{equation}
\begin{equation}\label{dzhdteqn}
\ddt{\zh} = 2 \frac{\epshe(\thy)}{\estarhe} -
\frac{(\zh-\zout)}{\sh}\left (2 \sdot-\ddt{\sh} \right ),
\end{equation}
\begin{eqnarray}\label{dthdteqn}
\ddt{\thy} = \frac{5}{4 \cp} \left [\estarh \gh \left (\frac{\zh +
\zout}{2}\right ) +\epshe(\thy) - \frac{\fhy - \fhe}{\sh} \right ] \nonumber \\ \nonumber \\ +
\frac{\thy}{4 \sh} \ddt{\sh},
\end{eqnarray}
\begin{equation}\label{dshedteqn}
\ddt{\she} = \left(2\sdot -\ddt{\sh}\right) - \frac{\delsig}{(1-\zh)}
\left [2 \frac{\epshe(\thel)}{\estarhe} -\ddt{\zh} \right ],
\end{equation}
\begin{eqnarray}\label{dthedteqn}
\ddt{\thel} = \frac{5 (\themthf)^{2}}{4\cp (\thel^{8} -5 \thel^{4}
\thy^{4} + 4 \thel^{3}\thy^{5})}
\times \nonumber \\ \nonumber \\
\left [\epshe(\thel) - \frac{\fhe}{\delsig} +
\frac{\cp(\thel-\thy)}{\delsig}\ddt{\sh}\right ] + 
\frac{\thel}{4\delsig}\left[1-\left(\frac{\thy}{\thel}\right)^{4}\right]
\ddt{\delsig}
\nonumber \\ \nonumber \\
 - \left (\frac{\thy^{8} -5 \thy^{4} \thel^{4} + 4
\thy^{3}\thel^{5}}{\thel^{8} -5\thel^{4} \thy^{4} + 4
\thel^{3}\thy^{5}}\right ) \ddt{\thy}. \nonumber \\
\end{eqnarray}
These are the fundamental equations of our two-zone model.  They can
be expressed in the compact form
\begin{equation}\label{compacteqn}
\ddt{\bf x} = {\bf f}_{\sdot,\xout,\zout}({\bf x}),
\end{equation}
where ${\bf x}$ represents the vector \{$\sh, \zh, \thy, \she,
\thel$\}, and $\sdot$, $\xout$, and $\zout$ are free parameters.

The reader should note that several of the dimensionless coefficients
in the above equations are $\sim 1$ and may be set to unity.
Furthermore, $\sh$ and $\she$ change over much longer timescales than
$\thy$ and $\thel$, and so the $d\sh/dt$ and $d\she/dt$ terms in
equations (\ref{dthdteqn}) and (\ref{dthedteqn}) can be set to zero.
Therefore, equations (\ref{dthdteqn}) and (\ref{dthedteqn}) can be
replaced by
\begin{equation}\label{dthdteqneasy}
\ddt{\thy} = \frac{5}{4\cp} \left [\estarh \gh \left (\frac{\zh +
\zout}{2}\right ) +\epshe(\thy) - \frac{\fhy - \fhe}{\sh} \right ],
\end{equation}
\begin{equation}\label{dthedteqneasy}
\ddt{\thel} = \frac{5}{4\cp} \left [\epshe(\thel) -
\frac{\fhe}{\delsig} \right ] - \ddt{\thy},
\end{equation}
with little loss of accuracy, although the results we present are for
the full equations (\ref{dshdteqn}-\ref{dthedteqn}).

\section{Equilibria and Their Stability}\label{eqandstab}

\begin{figure}
\plotone{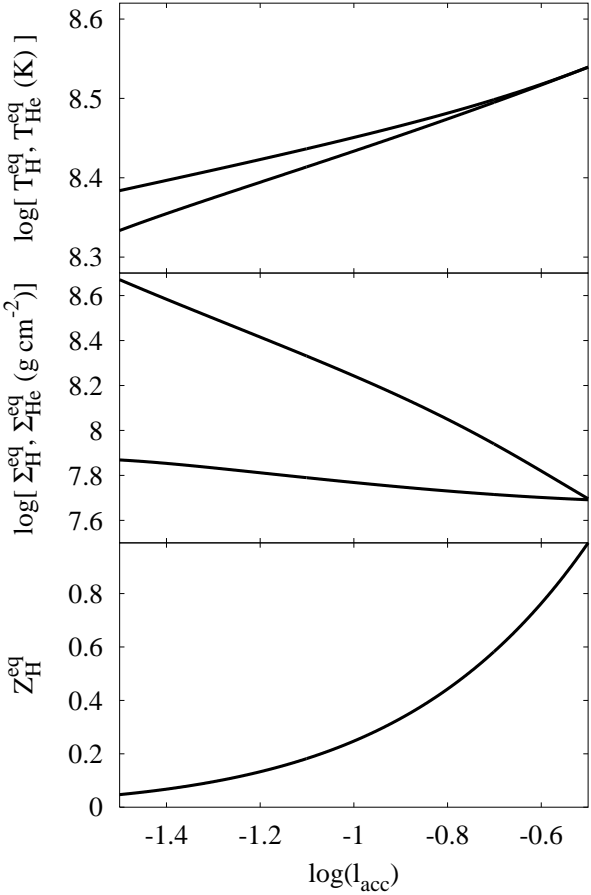}
\caption{Equilibrium values of the five variables $\thy$, $\thel$,
$\sh$, $\she$, and $\zh$ as a function of $\lacc \equiv \sdot /
\sdot_{\mathrm{Edd}}$.  Note that $\eq{\thel} > \eq{\thy}$ and
$\eq{\she} > \eq{\sh}$ for the indicated range of $\lacc$.  For larger
accretion rates, the inequalities are reversed and the model described
in this paper is no longer valid.}
\label{equilibria}
\end{figure}

Our goal in this section is to find the equilibrium solutions of the
above two-zone model and to determine their stability.  To find the
equilibrium solutions, we set $d/dt = 0$, so that equations
(\ref{dshdteqn}-\ref{dthedteqn}) become the following set of 5 coupled 
algebraic equations: 
\begin{equation}\label{eqeqn1}
\gh \left (\frac{\zh+\zout}{2} \right ) \sh = \sdot \xout,
\end{equation}
\begin{equation}
\epshe(\thy) \sh = \sdot (\zh-\zout)\estarhe,
\end{equation}
\begin{equation}
\fhy-\fhe = \estarh \gh \left (\frac{\zh+\zout}{2}
\right )\sh + \epshe(\thy)\sh,
\end{equation}
\begin{equation}
\epshe(\thel)  \delsig = \sdot (1-\zh)\estarhe,
\end{equation}
\begin{equation}
\fhe = \epshe(\thel)\delsig.
\end{equation}
We solve this set of equations for the 5 equilibrium values
$\eq{\sh}$, $\eq{\zh}$, $\eq{\thy}$, $\eq{\she}$, and $\eq{\thel}$.
When we do this, we obtain
\begin{equation}
\eq{\fhy} = \sdot \left[ \xout \estarh + (1-\zout)\estarhe \right ],
\end{equation}
\begin{equation}
\eq{\fhe} = \sdot (1-\eq{\zh})\estarhe.
\end{equation}
Thus the flux emitted from the stellar surface, $\eq{\fhy}$, equals
the flux released via steady-state nuclear burning of all the fuel,
and the flux entering zone (i), $\eq{\fhe}$, equals the flux released
via steady-state nuclear burning of the helium within zone (ii).  Note
that the equilibria are functions of $\sdot$, $\xout$, and $\zout$.
See Figure \ref{equilibria} for plots of the equilibrium values of the
five fundamental variables as a function of the Eddington-scaled
accretion rate
\begin{equation}
\lacc \equiv \frac{\sdot}{\sdot_{\mathrm{Edd}}},
\end{equation}
where $\sdot_{\mathrm{Edd}} = 1.0 \times 10^{5}$
$\mathrm{g}\,\mathrm{cm}^{-2}\,\mathrm{s}^{-1} $.  

The behavior of the equilibrium quantities shown in Figure
\ref{equilibria} is easily understood.  Nuclear burning generates
energy and heats the accreted layer.  In equilibrium, hydrogen and
helium burn at a rate $\propto \lacc$, and so the equilibrium
temperatures $\eq{\thy}$ and $\eq{\thel}$ are monotonically increasing
functions of $\lacc$.  The rate at which helium burns is very
temperature-sensitive, and so the depth at which helium depletes,
$\eq{\she}$, is a decreasing function of $\lacc$.  One-zone models
often presume that no stable helium burning occurs prior to a burst,
in which case $\eq{\zh} = \zout$.  This would imply that $\eq{\sh}
\propto \lacc$ from equation (\ref{eqeqn1}).  \citet{TP78} were the
first to remark that stable helium burning prior to ignition increases
the abundance of seed nuclei for the hot CNO cycle and thus expedites
hydrogen burning.  Figure \ref{equilibria} shows that this effect is
significant for the accretion rates we consider, since $\eq{\zh}$ is
often much greater than $\zout$.  Consequently, far from increasing
with $\lacc$, $\eq{\sh}$ is in fact a decreasing function of $\lacc$.

After we have solved for the equilibrium solution, we conduct a linear
stability analysis to determine whether the system is stable or
unstable to perturbations \citep[e.g.,][]{GH82}.  From equation
(\ref{compacteqn}),
\begin{equation}
\ddt{(\delta {\bf x})} = \left (\pd{{\bf f}}{{\bf x}}\right )_{{\bf x}
= \eq{{\bf x}}} (\delta {\bf x}),
\end{equation}
where $\partial {\bf f}/\partial {\bf x}$ denotes the Jacobian matrix,
which is evaluated at the equilibrium $\eq{{\bf x}}$.  We calculate
the eigenvalues of the Jacobian numerically using the standard
procedure for a real, nonsymmetric matrix \citep{WR71,PTVF92}.  If at
least one of the eigenvalues has a positive real part, then the system
is unstable to perturbations, and we say that type I X-ray bursts
occur.  Otherwise, the system is stable to bursts.  Thus the
eigenvalue with the greatest real part determines the bursting
behavior of the system.

\begin{figure}
\plotone{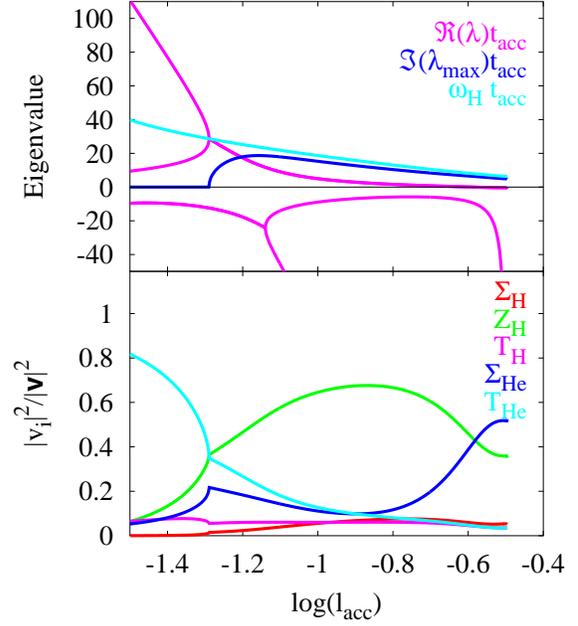}
\epsscale{1.2}
\caption{Top panel: spectrum of the real parts of the eigenvalues as a
function of $\lacc$.  Also shown are the imaginary part of the
eigenvalue with the greatest real part and $\omega_{\mathrm{H}} \equiv
2\pi/t_{\mathrm{H}}$, where $\tH \equiv \eq{\sh}/\sdot$ is the
hydrogen-burning timescale.  Each frequency is multiplied by the
accretion timescale $\tacc \equiv \eq{\she}/\sdot$.  Bottom panel:
normalized components of the eigenvector corresponding
to the eigenvalue with the greatest real part.}
\label{evalandvec}
\end{figure}

The top panel of Figure \ref{evalandvec} shows the spectrum of the
real parts of the eigenvalues $\lambda$ as a function of $\lacc$.  We
have multiplied the eigenvalues by the accretion timescale,
\begin{equation}
t_{\mathrm{acc}} \equiv \frac{\eq{\she}}{\sdot},
\end{equation}
to make them dimensionless.  The most negative eigenvalue has
$\reltacc \lesssim -1000$ and is not shown.  The bottom panel of
Figure \ref{evalandvec} shows the normalized squared moduli of the
components of the eigenvector ${\bf v}$ corresponding to the
eigenvalue with the greatest real part.  To construct ${\bf v}$, we
nondimensionalize the perturbations $\delta x_{i}$ such that the
components of the eigenvector are $v_{i} = \delta \ln (x_{i})$.

Figure \ref{evalandvec} nicely illustrates the nature of the
instability and the behavior of the resulting type I X-ray burst as a
function of the accretion rate.  For $\llacc \lesssim -1.3$, $\lambda$
is real and ${\bf v}$ is dominated by perturbations in $\thel$,
showing that a pure thermal perturbation in the helium burning region
triggers the instability.  Also, the perturbation grows very quickly
since $\reltacc \gg 1$.  Hydrogen burning is clearly inconsequential
to the instability since perturbations grow at a rate $\Re(\lambda) >
\oH$, where
\begin{equation}\label{oHeqn}
\oH \equiv \frac{2\pi}{\tH}
\end{equation}
and
\begin{equation}\label{tHeqn}
\tH \equiv \frac{\eq{\sh}}{\sdot}
\end{equation}
is the hydrogen-burning timescale.  These bursts (named helium
bursts by NH03) are triggered by the well-known and standard
thin-shell thermal instability.

For $-1.3 \lesssim \llacc \lesssim -0.9$, $\reltacc \gg 1$ so the
bursts are still ``prompt,'' but now $\lambda$ is complex.  This means
that there is an overstability rather than a pure instability.  Also,
$\Re(\lambda) < \oH$, which means that hydrogen burning now begins to
play a role in the stability of the system.  As a confirmation, we
note that ${\bf v}$ involves perturbations of all five parameters, in
particular the composition of the hydrogen-burning zone, showing that
all the variables play a role in the instability.

For $-0.9 \lesssim \llacc \lesssim -0.6$, $\reltacc \lesssim 1$ and
$\Im(\lambda)/\Re(\lambda) \gg 1$.  This is the regime of delayed
mixed bursts identified by NH03.  The growing mode is overstable and
undergoes several oscillations as the amplitude slowly increases.  The
angular frequency of the oscillations $\Im(\lambda) \approx \oH$,
showing that hydrogen burning plays a key role.  We have verified that
the imaginary parts of the eigenvalues in this regime derived using
the global linear stability analysis of NH03 are approximately equal
to $\oH$ as well.

Finally, all eigenvalues have negative real parts for $\llacc \gtrsim
-0.6$, and so no bursts occur at these high accretion rates.  The
$\she$ component begins to dominate ${\bf v}$ in this regime.  For
these systems, the effective radiative cooling rate of the accreted
layer, which is proportional to $\she^{-2}$ \citep[e.g.,][]{FHM81,B98},
is large enough to dampen any oscillations that might otherwise
initiate an instability.

It is now clear why we are able to drop the terms $\propto \sdot$ in
equation (\ref{entropyeqn}).  In the helium and prompt mixed burst
regimes, these terms are unimportant because thermal perturbations
drive the instability, and they occur on a timescale much shorter than
the accretion timescale $\tacc$ \citep[e.g.,][]{B98}.  In the
delayed mixed burst regime, although the mode grows on a timescale
comparable to the accretion timescale, compositional perturbations
primarily drive the overstability.

\begin{figure}
\epsscale{1.2}
\plotone{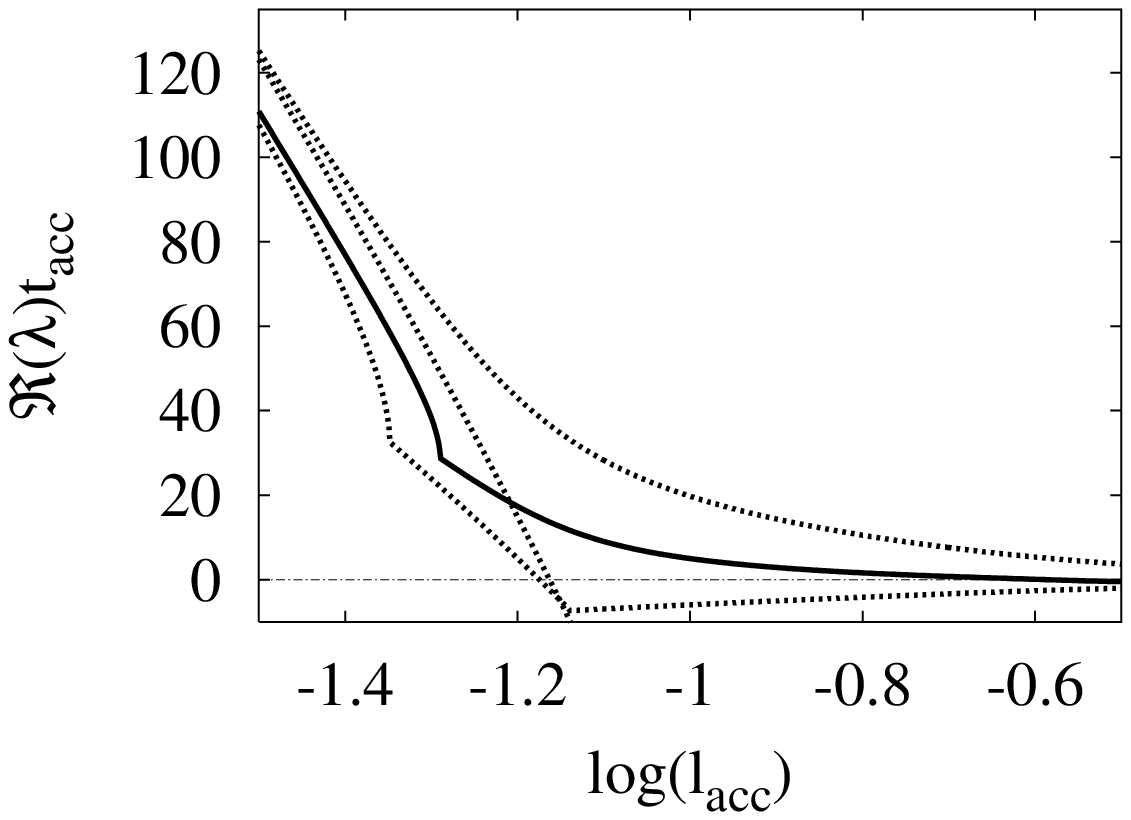}
\caption{Real part of the largest eigenvalue as a function of $\lacc$
for three models in which only a proper subset of the five parameters
$\sh$, $\zh$, $\thy$, $\she$, and $\thel$ is perturbed.  The dotted
lines show the normalized eigenvalues for models in which from top to
bottom (i) $\delta \sh = \delta \she = 0$, (ii) $\delta \sh = \delta
\she = \delta \zh = 0$, and (iii) $\delta \zh = 0$.  The solid line
shows the normalized eigenvalue for the full model.}
\label{subseteval}
\end{figure}

We consider our five-parameter, two-zone model to be ``minimal,'' by
which we mean that no model consisting of only a proper subset of the
five time-dependent variables $\sh$, $\zh$, $\thy$, $\she$, and
$\thel$ is able to reproduce the phenomenon of delayed mixed bursts.
To demonstrate this, and to illustrate the specific role each
parameter plays, we conduct linear stability analyses on truncated
$n$-parameter models, where $n<5$, and we compare the results of these
models to those of the full model.  To perform these calculations, we
determine the equilibria of the five parameters as usual, but we
perturb only the $n$ parameters $\delta x_{i}$ and set the
perturbations of the other $5-n$ parameters $\delta x_{i} = 0$.  We
then solve for the eigenvalues of the corresponding $n \times n$
Jacobian matrix.  See Figure \ref{subseteval} for plots of the real
part of the largest eigenvalue of each model as a function of $\lacc$.
The middle dotted line is the eigenvalue for an isobaric purely
thermal perturbation, where $\delta \sh = \delta \she = \delta \zh =
0$.  All equilibria of this model are stable for $\llacc \gtrsim
-1.2$.  This result nicely complements the eigenvector analysis shown
in Figure \ref{evalandvec}, which illustrates that thermal
perturbations drive the instability for $\llacc \lesssim -1.2$, but
they become less important for higher accretion rates.  Using their
global linear stability analysis, NH03 performed similar calculations
in which they considered only thermal perturbations, and they obtained
very similar results (see their Figures 16 and 17).  The bottommost
dotted line is the eigenvalue for a model in which only $\delta \zh =
0$.  This is a thermal perturbation as well, i.e., it includes no
perturbation in the nuclear composition parameter $\zh$, but it is one
in which the isobaric constraint is relaxed.  Figure \ref{subseteval}
shows again that all equilibria are stable for $\llacc \lesssim -1.2$.
However, the eigenvalues near the transition between stability and
instability are now complex.  This is evident from the kink in the
line at $\llacc \approx -1.35$, indicating that two real eigenvalues
have merged into a complex-conjugate pair.  The topmost dotted line is
the eigenvalue for a model in which $\delta \sh = \delta \she = 0$.
This perturbation is again isobaric, but it involves both thermal and
compositional perturbations.  Figure \ref{subseteval} shows that the
eigenvalue is real, and that all equilibria are unstable.

To better understand the role each parameter plays in the stability of
the dynamical system, we first divide the perturbations into two
classes.  Equations (\ref{dthdteqn}) and (\ref{dthedteqn}) imply that
$\delta \thy$ and $\delta \thel$ change on timescales on the order of
the thermal diffusion timescale, whereas equations (\ref{dshdteqn}),
(\ref{dshedteqn}), and (\ref{dzhdteqn}) imply that $\delta \sh$,
$\delta \she$, and $\delta \zh$ change on timescales on the order of
the accretion timescales $\tH$ and $\tacc$, which are
usually much longer than the thermal diffusion timescale.  Figures
\ref{evalandvec} and \ref{subseteval} demonstrate that, for $\llacc
\lesssim -1.2$, purely thermal perturbations drive the instability,
which grows on a timescale much shorter than the accretion timescale.
A positive temperature perturbations $\delta \thel > 0$ grows quickly
and accelerates helium burning, which decreases $\delta \she$.
Although a decrease in $\delta \she$ increases the effective cooling
rate, it does so over a much longer timescale, and so the effective
cooling rate is usually unable to compensate for the positive
temperature perturbation.  Therefore, the system is still thermally
unstable.  Thus, for thermal perturbations, it makes little difference
whether or not one allows $\delta \sh$ and $\delta \she$ to vary with
regard to the stability of the system \citep{P83,HCW05}.

The situation is quite different for the delayed mixed burst regime,
in which compositional perturbations $\delta \zh$ primarily drive the
instability.  A positive perturbation of any one of the three
parameters $\delta \thy$, $\delta \thel$, or $\delta \zh$ will cause
the other two to grow as well, for a positive temperature perturbation
accelerates helium burning, which produces more CNO and hence
increases $\delta \zh$, and a positive $\delta \zh$ perturbation
accelerates hydrogen burning, which raises the temperature and hence
increases $\delta \thy$ and $\delta \thel$.  Since this system is
stable to purely thermal perturbations, any unstable modes grow on the
slow accretion timescale.  But $\delta \sh$ and $\delta \she$, which
regulate the effective cooling rate, vary over a similar timescale.  
Therefore, $\delta \sh$ and $\delta \she$ have time to respond 
to changes in the nuclear heating rate in this regime.  A
positive perturbation in any one of $\delta \thy$, $\delta \thel$, or
$\delta \zh$ grows slowly and accelerates both hydrogen and helium
burning, which decrease $\delta \sh$ and $\delta \she$.  Decreases in
$\delta \sh$ and $\delta \she$ raise the effective cooling rate over a
similar timescale, which lowers the temperature and dampens nuclear
burning.  Consequently, $\delta \sh$ and $\delta \she$ then slowly
increase as freshly accreted matter advects into the layer until
nuclear burning begins again.  This cycle repeats, driving
oscillations with a period on the order of the hydrogen burning 
timescale $\tH$.

\section{Time Evolution of the Onset of Delayed Mixed Bursts}\label{tevolve}

Although linear stability analyses of the set of governing
differential equations clearly determine whether or not the system
will exhibit type I X-ray bursts, they can only hint at the nonlinear
development of the instability and the underlying physics of the burst
onset.  To remedy this, we numerically integrate equations
(\ref{dshdteqn}-\ref{dthedteqn}) to study the nonlinear evolution of
the onset of delayed mixed bursts.  We will not discuss the onset of
prompt bursts in any detail, since the physics of prompt bursts has
been previously studied by many authors and is generally well
understood
\citep{J78,TP79,JL80,Taam80,AJ82,HS82,T82,Pre83,TWWL93,Zetal01,Wetal04}.

\begin{figure}
\plotone{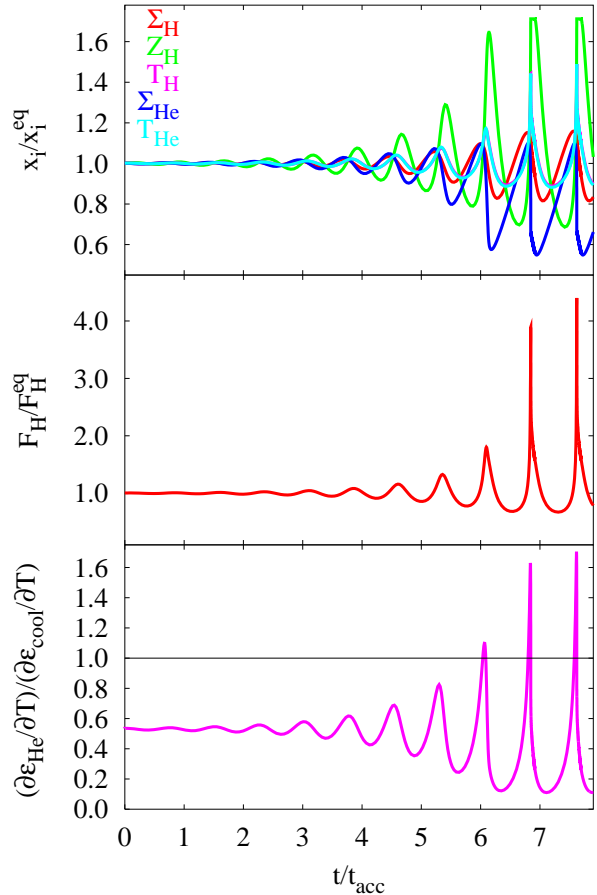}
\caption{Example of the time evolution of the overstability preceding
a delayed mixed burst.  $\xout = 0.7$, $\zout = 0.02$, and $\llacc
=-0.7$ for this calculation.  Top panel: evolution of the normalized
physical quantities.  Middle panel: normalized lightcurve. Bottom
panel: evolution of the ratio $(\partial \epshe/ \partial T)/(\partial
\epscool/ \partial T)$ evaluated at $\she$.  A thin-shell
thermonuclear instability ensues when $(\partial \epshe/ \partial
T)/(\partial \epscool/ \partial T) > 1$.}
\label{overstab}
\end{figure}

\begin{figure}
\plotone{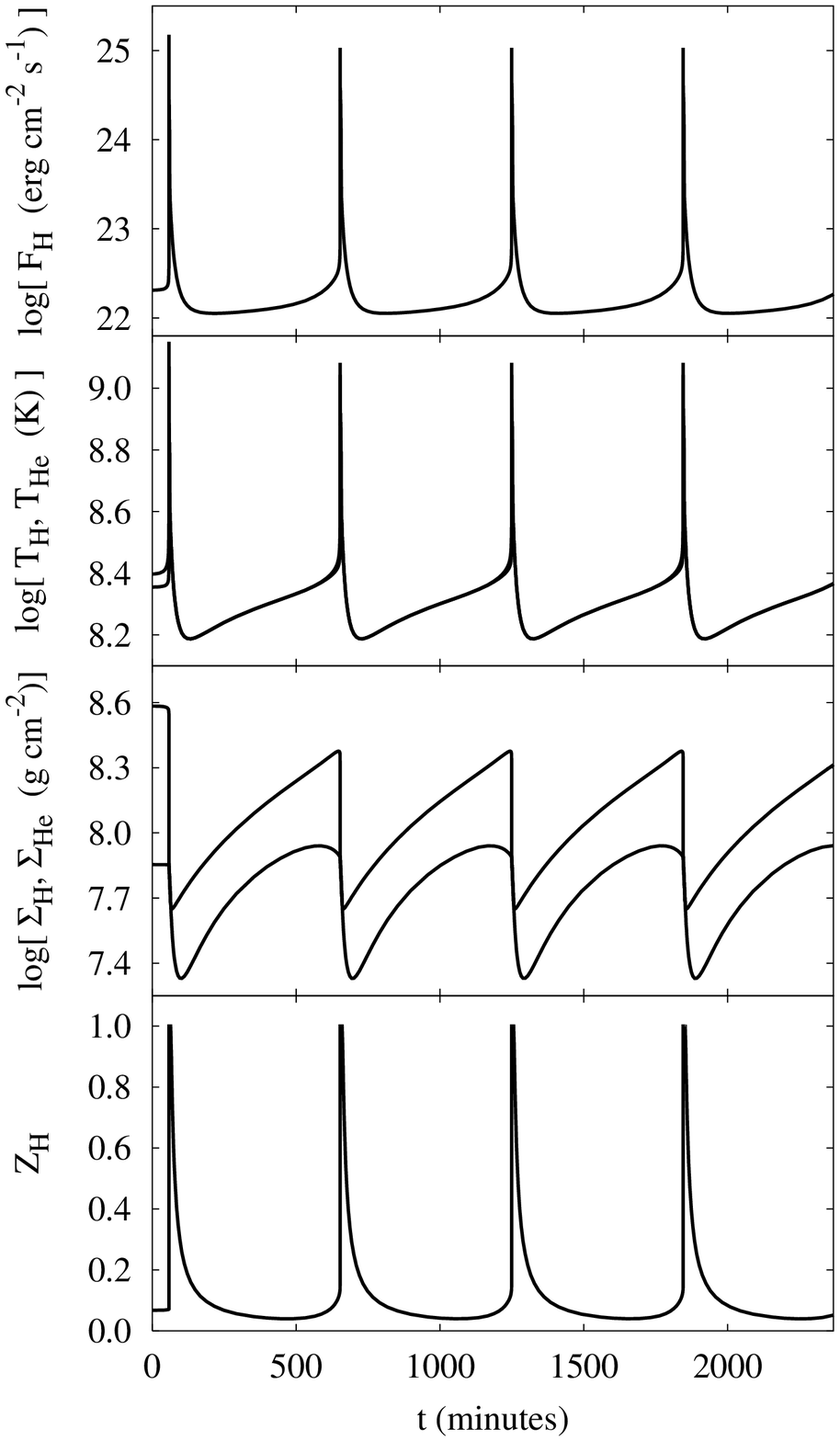}
\caption{Example of the time evolution of prompt bursts.  $\xout =
0.7$, $\zout = 0.02$, and $\llacc =-1.4$ for this calculation.  The
top panel shows the lightcurve, and the bottom three panels show the
time evolution of the five physical quantities.  Note that $\thel >
\thy$ and $\she > \sh$.}
\label{promptint}
\end{figure}

We perform the integration for a system with $\llacc = -0.7$, which is
well inside the delayed mixed burst regime.  Figure \ref{evalandvec}
shows that the system is marginally unstable, with $\reltacc \approx
0.75$, and that the system undergoes oscillations with a period
approximately equal to the hydrogen-burning timescale $\tH$.  We
initiate the overstability by perturbing the equilibrium such that the
initial conditions of the integration ${\bf x}_{0} = 1.001 \eq{\bf
x}$.  Figure \ref{overstab} shows the results of this calculation.
The top panel illustrates the time evolution of the physical
parameters.  All modes other than the principal mode are transient and
quickly decay after the initial perturbation.  $\thy$ and $\thel$ are
essentially in phase since the thermal diffusion timescale between
$\sh$ and $\she$ is much smaller than $\tH$.  The other physical
parameters are pairwise out of phase, but each slowly grows and
oscillates with the principal mode \citep{B93}.  One can ascertain the
relative phases by contrasting the phases of the various components of
the complex eigenvector ${\bf v}$.  For a comparison, we plot the time
evolution of the physical parameters for a sequence of prompt bursts
from a system with $\llacc = -1.4$ in Figure \ref{promptint}.  The
results of this prompt burst calculation are very similar to those of
previous one- and two-zone models \citep{BBL80,P83,RL84,HCW05}.

To better understand the physics of the overstability preceding a
delayed mixed burst, we consider a positive $\thy$ perturbation.
Since the thermal diffusion time from $\sh$ to $\she$ is very short,
$\thel$ follows the time dependence of $\thy$ throughout the
integration.  An increase in $\thy$ causes the helium in zone (i) to
burn at a higher rate, which increases $\zh$, and the simultaneous
increase in $\thel$ causes the helium in zone (ii) to burn at a higher
rate also, which decreases $\she$.  The larger $\zh$ causes hydrogen
to burn at a higher rate, which decreases $\sh$.  This raises the
effective cooling rate of the accreted layer, which is proportional to
$1/\sh^{2}$.  When nuclear burning has nearly depleted the hydrogen in
zone (i), both of the zones cool as freshly accreted matter advects
down to larger column depths.  $\thy$ and $\thel$ consequently
decrease until enough nuclear fuel has accumulated to restart nuclear
burning.  Once this happens, $\thy$ and $\thel$ will increase and the
cycle continues.

If the amplitudes of the oscillations increase in time, the
temperature $\thel$ at $\she$ may raise the helium nuclear energy
generation rate high enough to trigger a thin-shell thermal
instability and hence a type I X-ray burst.  To
understand when the burst will occur, we follow the basic one-zone,
single-parameter linear stability analysis of \citet{FHM81} and
consider the effective radiative cooling rate at $\she$
\begin{equation}\label{epscooleqn}
\epscool \equiv \frac{ac \thel^{4}}{3 \kappa \she^{2}}.
\end{equation}
A thin-shell thermal instability develops when
\begin{equation}\label{thinshelleqn}
\pd{\epshe}{T} > \pd{\epscool}{T}
\end{equation}
\citep{FHM81,HF82,FL87}.  We plot the ratio $(\partial \epshe/
\partial T)/(\partial \epscool/ \partial T)$ in the bottom panel of
Figure \ref{overstab}.  A comparison between the lightcurve shown in
the middle panel and the ratio $(\partial \epshe/ \partial
T)/(\partial \epscool/ \partial T)$ in the bottom panel shows that a
short type I X-ray burst, characterized by a sudden increase in the
outward flux, occurs when the thin-shell thermal instability criterion
$(\partial \epshe/ \partial T)/(\partial \epscool/ \partial T) > 1$ is
satisfied.  Note that the low peak flux of the burst is due to the
restriction in our model that hydrogen burns only via the
temperature-independent hot CNO cycle.  In our two-zone model, this
delayed mixed burst calculation evolves to a limit cycle of short
bursts with a recurrence time $\approx \tH$.  In reality, hydrogen
burning will proceed via the rp-process of \citet{WW81} during the
burst, and this will burn nearly all of the fuel and produce an
Eddington-limited flux \citep[e.g.,][]{HF84,Wetal04,FGWD06}.  The
system will then restart near $t/\tacc \approx 0$ in Figure
\ref{overstab} and undergo several oscillations before having its
next burst.  Our two-zone model, which includes no rp-process
reactions, is too simple to reproduce this behavior.  Note also that,
while equation (\ref{epscooleqn}) is the exact cooling rate for a
suitably constructed one-zone model, it is only approximate for our
two-zone model.  However, it is accurate enough to illustrate that,
much like prompt bursts, helium burning ultimately induces a
thin-shell thermal instability that triggers a delayed mixed burst.

\begin{figure}
\epsscale{1.2}
\plotone{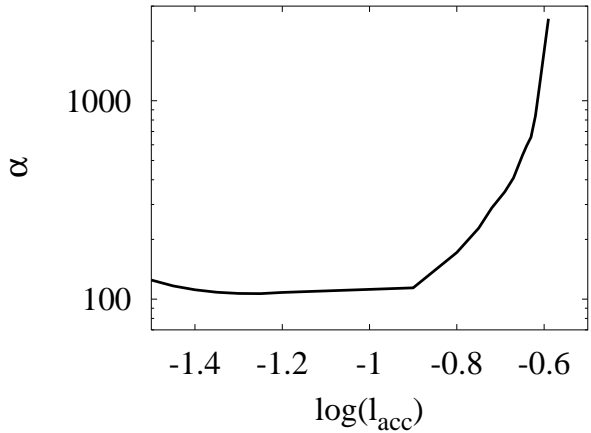}
\caption{Shows $\alpha$-values of type I X-ray bursts as 
a function of $\lacc$.  The steep rise in $\alpha$ beginning 
at $\llacc \approx -0.9$ as $\lacc$ increases is characteristic 
of delayed mixed bursts.}
\label{alphas}
\end{figure}

While the above calculation illustrates the physics of the onset of
delayed mixed bursts, it does not unambiguously demonstrate that the
bursts are in fact ``delayed.''  Specifically, we have not yet
demonstrated that a significant period of stable burning will precede
a type I X-ray burst.  Although we began the integration by perturbing
the equilibrium solution, a system that is dynamically unstable need
not come anywhere close to its equilibrium.  Unfortunately, we cannot
set the initial conditions to that of a bare neutron star, for which
${\bf x_{0}} =$ \{$\sh, \zh, \thy, \she, \thel$\} = \{$0, \zout, 0, 0,
0$\}, because the time steps required to begin the integration from
these initial conditions are too small to maintain numerical accuracy.
Instead, we set ${\bf x_{0}} =$ \{$\eq{\sh}, \eq{\zh}, \eq{\thy},
\eq{\sh}$, $\eq{\thy}$\}, i.e.~we start the system with the
hydrogen-burning layer in place but without any helium-burning
layer. We then integrate until a type I X-ray burst occurs.  This
calculation gives a recurrence time $t_{\mathrm{rec}} \equiv
\tH+t_{\mathrm{int}}$, where $t_{\mathrm{int}}$ is the integration
time.  From the recurrence time, we derive an estimate for the
dimensionless quantity $\alpha$, which is defined as the accretion
energy released between successive bursts divided by the nuclear
energy released during a burst.  Thus,
\begin{equation}\label{alphaeqn}
\alpha \approx \frac{\sdot c^{2} z t_{\mathrm{rec}}}{[\xout \estarh +
(1-\zout) \estarhe]\sh+ (1-\zh)\estarhe \she},
\end{equation}
where $z = (1-2GM/Rc^{2})^{-1/2} - 1$ is the gravitational redshift,
and the values of $\sh$, $\zh$, and $\she$ are determined from the
numerical integration just prior to the burst.  Both the accretion
energy and nuclear energy of the burst in equation (\ref{alphaeqn})
are those measured locally at the stellar surface.  In our expression
for $\alpha$, we implicitly assume that all of the hydrogen and helium
burns to CNO during a burst.  Figure \ref{alphas} shows that $\alpha
\sim 100$ in the regime of prompt bursts, which is consistent with
more sophisticated calculations \citep[e.g.,][]{Wetal04}, and that
$\alpha$ rises dramatically to values $\sim 1000$ for $\llacc \gtrsim
-0.9$, which illustrates that the systems in the delayed mixed burst
regime undergo long periods of stable nuclear burning prior to type I
X-ray bursts.  This is consistent with both the results of NH03 (see
their Figures 15 and 18) and observations \citep{vPPL88}.

\section{Discussion}\label{discussion}

In this section, we compare and contrast the results of the two-zone 
model presented in this work with the results of both the global linear 
stability analysis of NH03 and other theoretical models.  

\subsection{Comparison to the Global Stability Analysis of Narayan \& Heyl}

The primary goal of this work is to construct a simple model 
of nuclear burning on accreting neutron stars that elucidates the 
essential physics behind the more complicated and abstruse global 
linear stability analysis of NH03.  We assess how well we have 
accomplished this goal below.

We begin by discussing the various bursting regimes and the ranges of
accretion rates $\lacc$ in which these regimes lie.  A comparison
between the results of \S \ref{eqandstab} and Table 3 of NH03
demonstrates that the two-zone model reproduces all of the bursting
regimes of NH03 for the relatively high $\lacc$ considered in this
work, and that the ranges of $\lacc$ coincide very well with the
results of NH03's ``canonical'' neutron star, for which
$M=1.4M_{\odot}$, $R=10.4$ km, and $T_{\mathrm{core}} = 10^{8}$ K .
However, the two-zone model cannot reproduce the hydrogen bursts that
occur for $\llacc \lesssim -2.5$ because unstable thermonuclear
hydrogen burning triggers these burst, but our expression for $\epsh$
is temperature-independent.  We note that the $\lacc$ ranges of the
various regimes in the two-zone model depend somewhat on the values of
$\kappa$ and $\ghe$, assumed constant in our model.

The methods used to determine the equilibrium solution in each model
differ somewhat, although this difference should not affect the
eigenvalue.  NH03 started with an assumed column depth of accreted
plasma $\slayer$ and determined the equilibrium solution of the
physical quantities $\rho$, $T$, $F$, $X$, and $Y$ as functions of
$\Sigma$.  Next, they carried out a linear perturbation analysis to
determine the stability of the equilibrium solution.  If $\reltacc
\geq 3$, where in this case $\tacc \equiv \slayer/\sdot$, then the
configuration was unstable and a type I X-ray burst was assumed to
occur.  If $\reltacc < 3$, the configuration was stable, and the
process was repeated for a larger value of $\slayer$.  If $\reltacc <
3$ for all trial values of $\slayer$, then the system was stable, and
no type I X-ray bursts occurred.  It is important to understand that
the method of NH03 will find all systems that have positive
eigenvalues to be unstable because $\tacc$ increases as $\slayer$
increases.  In the two-zone model, we assume complete burning of the
accreted matter, and we say that type I X-ray bursts occur if
$\reltacc > 0$ and that no bursts occur otherwise.  If a system is
stable to bursts, the accreted matter steadily burns to completion,
and both methods agree that this equilibrium is stable.  If a system
is unstable to bursts, the two-zone model determines that the complete
steady-burning equilibrium solution is unstable.  The method of NH03
may find that the identical system is unstable either before all of
the fuel burns to completion (in a prompt burst, for example) or after
all the fuel burns to completion (in, say, a delayed burst), but it
will always determine that the system is indeed unstable.  Therefore,
although the two stability analyses differ to some extent, these
differences are unimportant with respect to the determination of the
eigenvalue of a system.

The eigenvalues derived using the two models agree quite well.  We
have shown already that the real parts of the eigenvalues agree quite
well, for they delineate between the prompt and delayed bursting
regimes.  We now compare the imaginary parts.  We have shown in \S
\ref{eqandstab} that $\Im(\lambda) \approx \oH$ in the delayed mixed
burst regime, which means that oscillations occur with a period equal
to the hydrogen-burning timescale.  To explicitly show that
$\Im(\lambda) \approx \oH$ in the global linear stability analysis, we
consider the calculation presented in \S 4.2 of NH03.  In that
calculation, NH03 consider a system with $\llacc = -0.9$, which
implies that $\sdot = 1.04 \times 10^{4}$
$\mathrm{g}\,\mathrm{cm}^{-2}\,\mathrm{s}^{-1}$, since
$\sdot_{\mathrm{Edd}} = 8.26 \times 10^{4}$
$\mathrm{g}\,\mathrm{cm}^{-2}\,\mathrm{s}^{-1}$ for that system.
Figures 8 and 9 of NH03 imply that $\eq{\sh} \approx 1.0 \times
10^{8}$ $\mathrm{g}\,\mathrm{cm}^{-2}$, and so $\oH = 6.5 \times
10^{-4}$ $\mathrm{s}^{-1}$ from equations (\ref{oHeqn}) and
(\ref{tHeqn}).  This value of $\oH$ is very close to the $6.6 \times
10^{-4}$ $\mathrm{s}^{-1}$ NH03 quote for the imaginary part of the 
eigenvalue they obtained in their calculation.

\citet{RCGS01} discovered a class of low-frequency oscillations that
precede type I X-ray bursts in the systems 4U 1608$-$522, 4U
1636$-$536, and Aql X-1.  These oscillations have periods $\sim 120$
seconds and have been observed when the accretion rate $\lacc \approx
0.1$.  While the modes resulting from the global stability analysis of
NH03 are complex in this regime, the oscillation periods of these
modes are roughly a factor of $10$ larger than the observed periods.
It is now clear why NH03 were unsuccessful in explaining these
observations.  We have shown in \S \ref{eqandstab} that the
oscillation periods of overstable modes in this regime are
approximately equal to the hydrogen-burning timescale $\tH$.  From
equations (\ref{eqeqn1}) and (\ref{tHeqn}),
\begin{equation}
\tH = \frac{2 \xout}{\gh (\eq{\zh}+\zout)} > 770 \left (\frac{\xout}{0.7}
\right ) \mathrm{s}.
\end{equation}
Thus the oscillations preceding a delayed mixed burst have too long of
a period and cannot be associated with the oscillations observed by
\citet{RCGS01} if hydrogen burns predominantly via the hot CNO cycle.

We have shown in \S 4 that, in the delayed mixed burst regime, a
considerable amount of stable burning occurs prior to a type I X-ray
burst.  Consequently, $\alpha$, the ratio of the gravitational energy
released via accretion to the nuclear energy released during a burst,
rises dramatically near the critical $\lacc$ above which bursts do not
occur.  A comparison between Figure \ref{alphas} of this work and
Figure 15 of NH03 demonstrates that the $\alpha$-values derived using
the two different models agree quite well.

\begin{figure*}
\plottwo{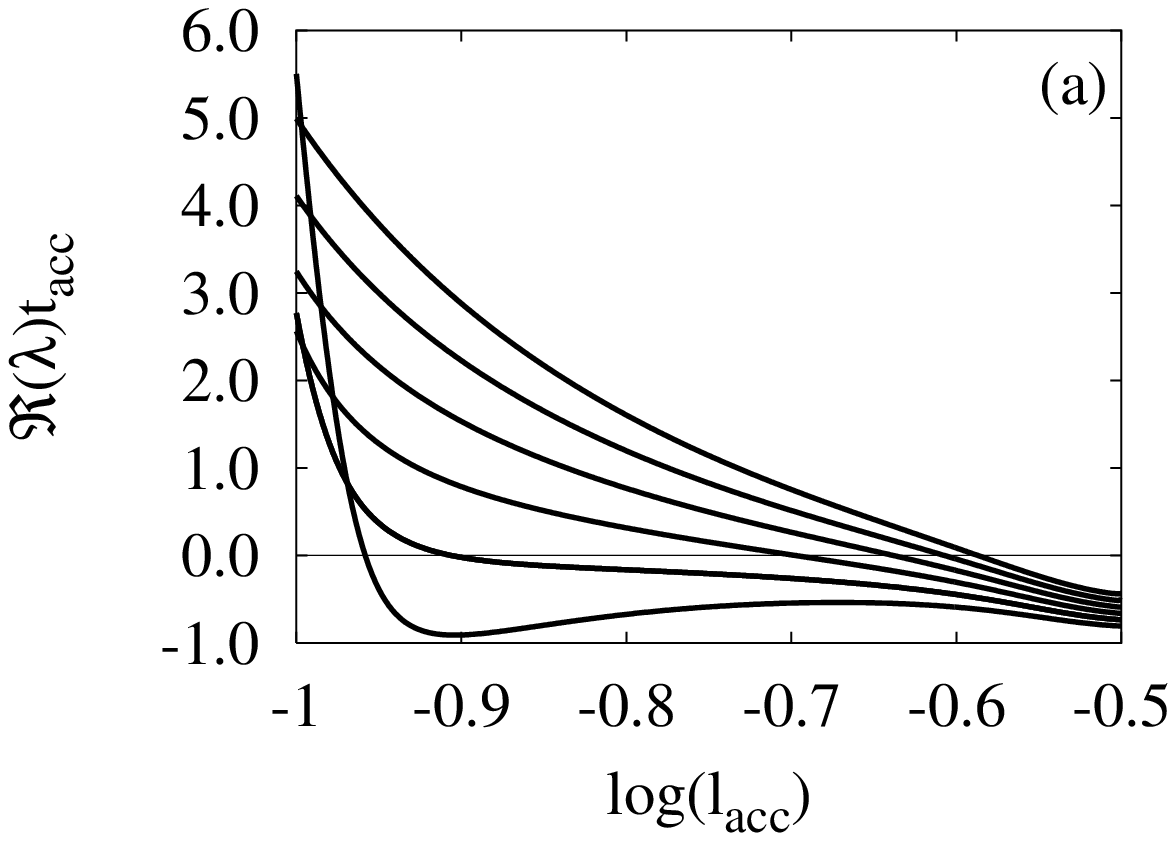}{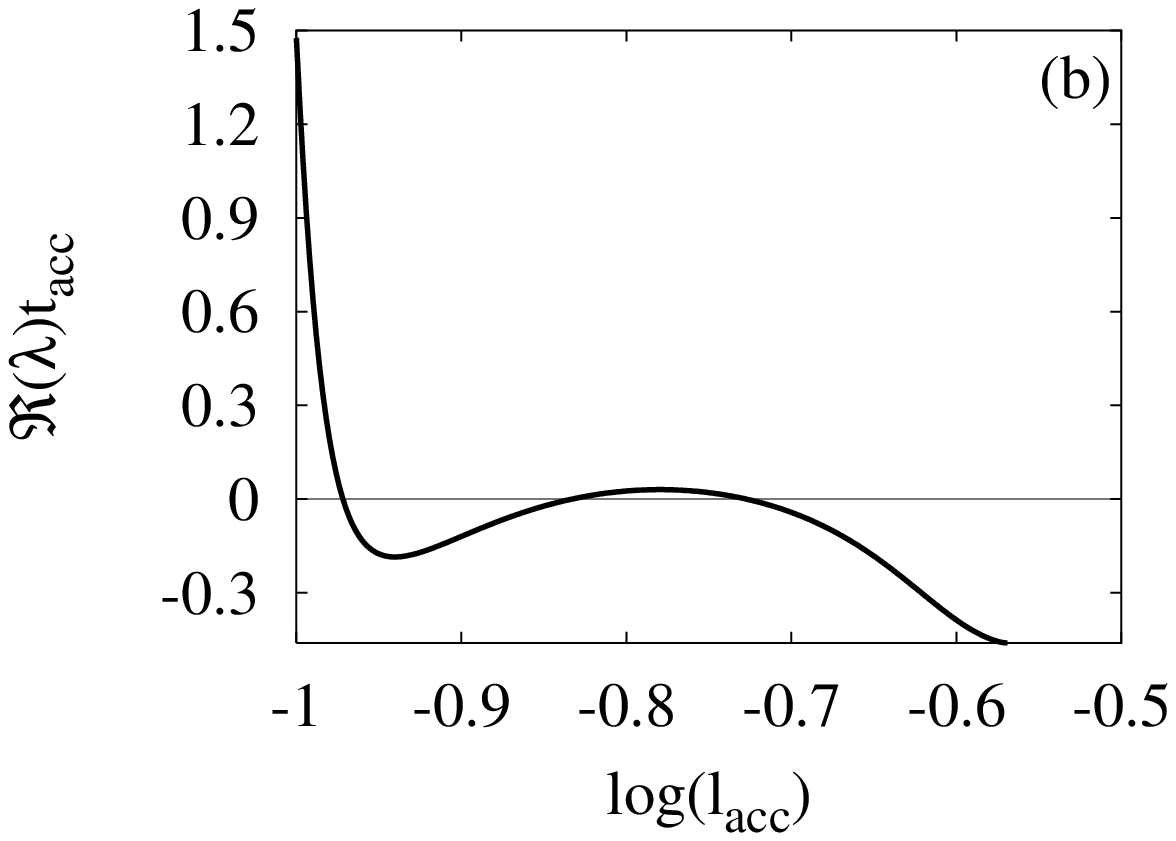}
\caption{Real part of the largest eigenvalue as a function of $\lacc$.
($a$) Normalized eigenvalue for models in which from top to bottom
$\zout = 0.02$, $0.04$, $0.06$, $0.08$, $0.10$, and $0.12$.  ($b$)
Normalized eigenvalue for a model in which $\zout = 0.132$ and $\ghe =
0.3$.  The prompt mixed burst and delayed mixed burst regimes 
are clearly separated by a regime of stable nuclear burning.  }
\label{Zgraphs}
\end{figure*}

Using an improved version of the model of NH03, \citet{CMSN06} found
that increasing $\zout$, the CNO abundance of the accreted plasma,
diminishes the range of $\lacc$ over which delayed mixed bursts occur.
In fact, for large enough values of $\zout$, \citet{CMSN06} found that
the ranges of $\lacc$ in which prompt mixed bursts and delayed mixed
bursts occur are separated by a regime of stable burning.  Figure
\ref{Zgraphs}$a$ shows the normalized eigenvalues of the two-zone
model for systems in which $\zout \geq 0.02$.  The critical $\lacc$
above which bursts do not occur decreases with increasing $\zout$, in
agreement with Figure 5 of \citet{CMSN06}, but the two-zone model does
not reproduce the regime of stable burning that separates the prompt
mixed bursts from the delayed mixed bursts for our choices of $\kappa$
and $\ghe$.  However, we are able to reproduce this regime with
certain combinations of $\zout$, $\kappa$, and $\ghe$.  Figure
\ref{Zgraphs}$b$ shows the normalized eigenvalues for a model in which
$\zout = 0.132$ and $\ghe = 0.3$.  A stable burning regime separating
prompt mixed bursts from the delayed mixed bursts is clearly evident
for this calculation, showing that the two-zone model has the right
qualitative behavior.  However, our choice of $\zout$ for this 
calculation is more than twice that required in the more accurate model of
\citet{CMSN06}, showing that (not surprisingly) there are quantitative
discrepancies.

\subsection{Comparison to Previous Theoretical Work}

For accretion rates $\lacc \lesssim 0.1$, type I X-ray bursts are
well-understood theoretically, and the results of theoretical models
are generally in accord with observations.  This is not true for
$\lacc \gtrsim 0.1$.  Nearly all theoretical models predict that
accreting neutron stars should exhibit type I X-ray bursts for all
$\lacc \lesssim 1$ \citep{FHM81,AJ82,T85,TWL96,B98,FHLT03,HCW05},
whereas the model of NH03 and the two-zone model presented in this
work suggest that bursts should cease for $\lacc \gtrsim 0.25$, in
agreement with observations \citep{vPCLJ79,vPPL88,Cetal03,RLCN06}.
Furthermore, our model and that of NH03 are the only models able to
explain the dramatic increase in $\alpha$ with increasing $\lacc$
observed by \citet{vPPL88}.  In this section, we compare and contrast
the two-zone model presented in this work with previous one-zone burst
models.  Additionally, we discuss the results of this work in relation
to those of detailed time-dependent multi-zone models that utilize
sophisticated nuclear reaction networks.

We begin by discussing the stability criterion of one-zone models
\citep[e.g.,][]{FHM81,P83,B98,CB00,HCW05}.  These models presume that
helium burning (or hydrogen burning for very low $\lacc$) initiates a
thermonuclear instability and produces a type I X-ray burst.  As
mentioned in \S \ref{themodel}, the hydrogen burning rate is
temperature-independent for $T \gtrsim 10^{8}$ K and so hydrogen
burning cannot be thermally unstable for $\lacc \gtrsim 0.01$.  Since
all of the one-zone models give very similar results, we focus on the
model of \citet{HCW05}, which is an extension of the model of
\citet{P83}, for a comparison.  Written in our notation, the
differential equations (1-2) of \citet{HCW05} that govern their model
are
\begin{equation}\label{HCWeq1}
\ddt{\she} = \sdot - \she \frac{\epshe(\thel)}{\estarhe},
\end{equation}
\begin{equation}\label{HCWeq2}
\ddt{\thel} = \frac{1}{\cp}\left [\epshe(\thel) - \frac{\fhe}{\she}
\right],
\end{equation}
where $\fhe = ac \thel^{4}/3 \kappa \she$.  They presume that
$\eq{\sh} > \eq{\she}$ and that no stable helium burning occurs prior
to the burst.  The equilibrium solution to equations
(\ref{HCWeq1}-\ref{HCWeq2}) is
\begin{equation}\label{HCWeq}
\epshe(\thel) \she = \fhe = \sdot \estarhe,
\end{equation}
or equivalently,
\begin{equation}\label{he=cool}
\epshe(\thel) = \epscool,
\end{equation}
where $\epscool$ is given by equation (\ref{epscooleqn}).  Using the
thin-shell thermal instability criterion of equation
(\ref{thinshelleqn}), one finds that the transition from stability to
instability occurs when $\nu \equiv \partial \ln \epshe / \partial \ln
T \approx 4$, since $\partial \ln \epscool / \partial \ln T = 4$.  One
finds nearly the same condition by performing a standard linear
stability analysis on equations (\ref{HCWeq1}) and (\ref{HCWeq2}).
Thus, from equation (\ref{epsheeqn}), the equilibria are stable to
bursts when $\eq{\thel} \gtrsim 6 \times 10^{8}$ K.  However, Figures
\ref{equilibria} and \ref{subseteval} illustrate that many of the
equilibria in our two-zone model are stable to thermal perturbations
for temperatures much less than $6 \times 10^{8}$ K.  To understand
this, note that equation (\ref{he=cool}) implies that helium burning
is the only source of nuclear energy generation.  In our two-zone
model, both hydrogen burning and helium burning are sources of nuclear
energy generation, and so qualitatively $\epscool \sim \epshe(T) +
\epsh(Z)$.  Hydrogen burning is temperature-independent, so equation
(\ref{thinshelleqn}) still gives the right criterion for a thin-shell
thermal instability.  However, because $\epscool > \epshe$, the
critical value of $\nu$ for instability is larger than $4$.  Roughly,
on expects something like $\nu_{\mathrm{crit}} \sim
4(\epscool/\epshe)$.  Several authors \citep[e.g.,][]{FHM81,CB00} have
included the effect of hydrogen burning on the stability criterion in
their one-zone models.  These models assume that no helium burning
takes place prior to ignition, so $Z = \zout$ throughout the accreted
layer.  Consequently, they find that the effect of hydrogen burning on
the stability of the system is minor.  In contrast, stable helium
burning prior to ignition is substantial for $\lacc \gtrsim 0.1$ in
both our two-zone model and that of NH03, and so the effect of
hydrogen burning on the stability of the system is considerable.
Thus, hydrogen burning via the temperature-independent hot CNO cycle,
augmented by the extra CNO produced from stable helium burning, helps
stabilize nuclear burning on accreting neutron stars even at
temperatures $\sim 3 \times 10^{8}$ K, which is much less than the
$\sim 6 \times 10^{8}$ K needed to stabilize pure helium burning.

In the limit $\xout \rightarrow 0$ and $\zout \rightarrow 0$,
equations (\ref{dshdteqn}-\ref{dthedteqn}) of our two-zone model
reduce to the following set of two coupled differential equations:
\begin{equation}\label{limiteqn1}
\ddt{\she} = 2 \left [\sdot - \she \frac{\epshe(\thel)}{\estarhe}
\right],
\end{equation}
\begin{equation}\label{limiteqn2}
\ddt{\thel} = \frac{5}{4\cp}\left [\epshe(\thel) - \frac{\fhe}{\she}
\right] + \frac{\thel}{4 \she} \ddt{\she},
\end{equation}
where again $\fhe=ac \thel^{4}/3 \kappa \she$ from equation
(\ref{fheeqn}), and the equilibrium is given by equation
(\ref{HCWeq}).  Equations (\ref{limiteqn1}) and (\ref{limiteqn2}) are
effectively the same as equations (\ref{HCWeq1}) and (\ref{HCWeq2}).
With the system of equations (\ref{limiteqn1}-\ref{limiteqn2}), we are
able to reproduce the results of the one-zone models of \citet{P83}
and \citet{HCW05}.  As expected, the critical temperature 
for stability is $\sim 6 \times 10^{8}$ K in this case.

It is perhaps not surprising that one-zone models that focus
only on helium burning are not able to reproduce all of the bursting
regimes found using global linear stability analyses or observed in
nature.  As we have shown in this paper, one needs at least two zones,
one in which only helium burns and another in which both hydrogen and
helium burn together, to incorporate all of the key physics included 
in the
model of NH03.  One-zone models simply do not include enough physics.
More troubling is the fact that the results of detailed multi-zone
calculations of type I X-ray bursts
\citep[e.g.,][]{AJ82,TWL96,FHLT03,HCW05} do not reproduce the regime of
delayed mixed bursts and are also inconsistent with observations for
$\lacc \gtrsim 0.1$.  Since we do not have access to such models, we
are unable to conduct a direct comparison between our model and these
numerical calculations.  However, we attempt to better understand this
discrepancy by considering the work of \citet{FGWD06}.  

\citet{FGWD06} study the effects of the relatively unconstrained
$^{15}$O($\alpha$,$\gamma$)$^{19}$Ne reaction rate on type I X-ray
bursts from neutron stars accreting at $\lacc \approx 0.1$.  This
reaction is one of the two hot CNO cycle breakout reactions
\citep{W69,WW81,LWFG86,WGS99,SBCW99}.  For the lowest of the three trial
reaction rates they consider, they find that, after an
initial type I X-ray burst, hydrogen and helium burn stably via the
hot CNO cycle and triple-$\alpha$ reactions, respectively, and the
nuclear burning generates a slowly oscillating luminosity with a
period approximately equal to the hydrogen-burning timescale.  The
nuclear burning behavior in this calculation is nearly identical to
that of the nuclear burning preceding a delayed mixed burst found in
\S \ref{tevolve}.  This similarity is perhaps to be expected, for the
limited reaction networks employed in the model of NH03 and the
present model omit hot CNO cycle breakout reactions entirely.  For the two
higher trial reaction rates investigated by \citet{FGWD06}, the
breakout sequence
$^{15}$O($\alpha$,$\gamma$)$^{19}$Ne($p$,$\gamma$)$^{20}$Na diminishes
the CNO abundance and thus reduces the rate of hydrogen burning.
Hence, when helium ignition commences, the helium energy generation
rate dominates the total nuclear energy generation rate, and a prompt
type I X-ray burst occurs via a thin-shell thermal instability.  We
tentatively suggest that the discrepancies between the results of the
global linear stability analysis of NH03 and the results of multi-zone
calculations arise from differences in the treatment of hot CNO cycle 
breakout reactions.  In the NH03 model and our two-zone model, 
hydrogen burning via the
temperature-independent hot CNO cycle helps stabilize nuclear burning
on accreting neutron stars.  Breakout reactions would reduce the
degree to which hydrogen burns via the hot CNO cycle and thereby
increase the temperature-sensitivity of the total effective nuclear
energy generation rate.  Thus, breakout reactions would presumably
extend the range of accretion rates in which nuclear burning is
thermally unstable, but we do not include these reactions.  
That observations agree much better with the
results of NH03 and our two-zone model may imply that the true cross
sections of reactions such as the hot CNO cycle breakout reactions
$^{15}$O($\alpha$,$\gamma$)$^{19}$Ne and
$^{18}$Ne($\alpha$,$p$)$^{21}$Na are much smaller than the cross
sections employed in the reaction networks of multi-zone models, i.e.,
the true $^{15}$O($\alpha$,$\gamma$)$^{19}$Ne reaction rate is closer
to the lowest reaction rate considered by \citet{FGWD06}.  By omitting
hot CNO cycle breakout reactions altogether, the model of NH03 is
perhaps a better representation of the nuclear burning that precedes
type I X-ray bursts than time-dependent multi-zone models as 
presently implemented.

The model we present in \S \ref{themodel} is valid only when $\sh <
\she$, which restricts the range of accretion rates we can study to
$\lacc \lesssim 0.3$.  We have constructed a similar model to
determine the stability of nuclear burning on accreting neutron stars
for which $\sh > \she$, allowing us to study the stability of nuclear
burning at higher $\lacc$.  Although we find that all equilibria for
$\lacc \gtrsim 0.3$ are stable, which is consistent with the results
of NH03, we cannot state with confidence that either the two-zone
model or the model of NH03 is an accurate representation of nuclear
burning on accreting neutron stars for such high accretion rates.  For
$\lacc \gtrsim 1$, nuclear reactions other than the hot CNO cycle and
triple-$\alpha$ reaction almost certainly play a significant role
in the nuclear burning prior to a type I X-ray burst.  It is possible
that there exists another unstable burning regime near $\lacc \sim 1$,
and the range of $\lacc$ in which this regime might exist would
probably depend upon the $\lacc$ at which the effective
hydrogen-burning rate becomes predominantly temperature-dependent
rather than composition-dependent.  While most low-mass X-ray binaries
with $\lacc \gtrsim 0.3$ do not have type I X-ray bursts, GX 17+2 and
Cyg X-2 are notable exceptions
\citep{KG84,THKMM84,Setal86,KvdKvP95,Ketal97,Wetal97,S98,KHvdKLM02},
although these sources may exhibit type I X-ray bursts for other
reasons.  For example, bursts could occur at $\lacc \sim 1$ if the
accreted plasma is hydrogen-deficient \citep[e.g.,][]{CMSN06}.

\section{Conclusions}\label{conclusions}

We have constructed a simple two-zone model of type I X-ray bursts on
accreting neutron stars.  This model reproduces the delayed mixed
burst regime of NH03 as well as the helium and prompt mixed
burst regimes of previous studies
(Fujimoto et~al.\ 1981; Fushiki \& Lamb 1987; Cumming \& Bildsten
2000; NH03), and it agrees well with observations of type I X-ray
bursts \citep{vPCLJ79,vPPL88,Cetal03,RLCN06}.  More importantly, the model
illustrates the physics of the onset of instability as a function of
the local accretion rate $\sdot$, and it facilitates comparisons between
global linear stability analyses (which are more accurate but
difficult to understand physically) and other burst models.

A pure, rapidly growing thermal instability in the helium-burning zone
triggers bursts at relatively low $\sdot$.  As $\sdot$ increases above 
$0.1 \sdot_{\mathrm{Edd}}$, the
trigger mechanism evolves from that of a thermal instability to that
of a slowly growing overstability involving all parameters,
particularly the CNO mass fraction $\zh$ of the hydrogen-burning zone.
The competition between nuclear heating via the $\beta$-limited CNO
cycle as well as the triple-$\alpha$ process and radiative cooling via
outward diffusion of photons coupled with radiation from the stellar
surface drives oscillations with a period approximately equal to the
hydrogen-burning timescale $\eq{\sh}/\sdot$.  If these oscillations
grow in time, the temperature $\thel$ at the base of the helium layer
will rise, eventually triggering a thin-shell thermal instability and
hence a delayed mixed burst.  For $\sdot/\sdot_{\mathrm{Edd}} \gtrsim
0.25$ radiative cooling from the stellar surface dampens the
overstability, and no bursts occur.  We consider our two-zone model to
be ``minimal,'' by which we mean that no model consisting of only a
proper subset of the five time-dependent variables $\sh$, $\zh$,
$\thy$, $\she$, and $\thel$ is able to reproduce the phenomenon of
delayed mixed bursts.

Nearly all other theoretical models predict that bursts should occur
for all $\dot{\Sigma}/\dot{\Sigma}_{\mathrm{Edd}} \lesssim 1$, in
disagreement with the results of both NH03 and the two-zone model, as
well as with observations.  We suggest that this discrepancy arises
from the assumed strength of the hot CNO cycle breakout reaction
$^{15}$O($\alpha$,$\gamma$)$^{19}$Ne \citep{W69,WW81,SBCW99,FGWD06} in
time-dependent multi-zone burst models.  That
observations agree much better with the results of both NH03 and this
work may imply that the true $^{15}$O($\alpha$,$\gamma$)$^{19}$Ne
cross section is much smaller than the cross sections employed in the
reaction networks of these models.  Further calculations
such as those presented by \citet{FGWD06} in which the
$^{15}$O($\alpha$,$\gamma$)$^{19}$Ne reaction rate in the networks of
time-dependent multi-zone models is varied should be performed.

We have considered only two forms of nuclear burning in our model:
hydrogen burning via the hot CNO cycle and helium burning via the
triple-$\alpha$ process.  While this simplification is almost surely
reasonable for studying the onset of helium bursts, certain additional
reactions may be important in the delayed mixed burst regime or at
yet higher $\sdot$.  In particular, hot CNO cycle breakout reactions
such as $^{15}$O($\alpha$,$\gamma$)$^{19}$Ne could significantly
affect the CNO metallicity $\zh$ of the hydrogen burning zone and
consequently have some effect on delayed mixed bursts.  Additional
hydrogen burning processes that could circumvent the hot CNO cycle
such as the $^{15}$O($p$,$\gamma$)($\beta^{+}
\nu$)$^{16}$O($p$,$\gamma$)$^{17}$F($p$,$\gamma$)$^{18}$Ne($\beta^{+}
\nu$)$^{18}$F($p$,$\alpha$)$^{15}$O reaction sequence \citep{Setal06}
could affect delayed mixed bursts as well.  These issues are worthy of
further investigation.

\acknowledgments

It is a pleasure to thank Lars Bildsten, Edward Brown, Andrew Cumming,
and Sanjib Gupta for helpful discussions and the referee for comments
and suggestions that significantly improved the quality and utility of
this investigation.  This work was supported by NASA grant NNG04GL38G.


\clearpage

\end{document}